\documentclass{iopjournal}

\usepackage{amsmath}
\usepackage{amssymb}
\usepackage{bm}
\usepackage{booktabs}
\usepackage{threeparttable}
\usepackage{appendix}
\usepackage[numbers,sort&compress]{natbib}
\newtheorem{theorem}{Theorem}

\begin{document}

\articletype{Paper} 

\title{Discovery of connectivity-trainability trade-off of IQP Circuits for Hamiltonian optimization}

\author{Quoc Chuong Nguyen$^1$\orcid{0000-0002-3260-9967}, Dao Duy Tung$^{2,4}$ and Vu Tuan Hai$^{3,4,*}$\orcid{0000-0002-2293-0968}}

\affil{$^1$Institute of Fundamental and Applied Sciences, Duy Tan University, Ho Chi Minh City, 700000, Vietnam}

\affil{$^2$University of Science, Ho Chi Minh City, 700000, Vietnam}

\affil{$^3$University of Information Technology, Ho Chi Minh City, 700000, Vietnam}

\affil{$^4$Vietnam National University, Ho Chi Minh City, 700000, Vietnam}

\affil{$^*$\email{haivt@uit.edu.vn}}


\begin{abstract}
Instantaneous Quantum Polynomial-time (IQP) circuits are promising candidates for near-term quantum computing due to the conjectured classical hardness of their sampling task and their shallow, highly parallelizable structure. Despite extensive studies of their complexity-theoretic properties, the role of circuit connectivity in their trainability as variational ansatze remains poorly understood. In this work, we present an investigation of IQP circuits for Hamiltonian optimization that traces a mechanistic chain linking circuit connectivity, entanglement generation, and gradient scaling: increasing connectivity drives the circuit output toward Haar-like statistics and volume-law entanglement, and this increase in entanglement is consistently associated with an exponential decay of the gradient variance and degraded optimization performance, while sparser architectures retain stable gradients at the cost of global expressibility and entanglement capacity. This connectivity-entanglement-gradient mechanism produces a nontrivial connectivity-trainability trade-off, demonstrating that effective IQP-based Hamiltonian optimization requires a balance between global expressibility and trainability rather than maximal expressivity alone. In particular, on our benchmarks, a circuit generating no entanglement already matches the optimization performance of more highly connected circuits, and a circular topology retains entanglement capacity while preserving trainability, making it a broadly effective choice among the architectures considered.
\end{abstract}

\section{Introduction}

Combinatorial optimization problems arise in a broad range of scientific and engineering applications, including scheduling, portfolio optimization, logistics, machine learning, and computational physics. Many of these problems are NP-hard and can be reformulated as the task of finding the ground state of a suitably defined Hamiltonian \cite{10.3389/fphy.2014.00005}. Such Hamiltonian optimization problems establish a natural connection between optimization and quantum computing, motivating the development of quantum algorithms capable of preparing low-energy states efficiently. In particular, the variational quantum eigensolver (VQE) framework has emerged as one of the most promising paradigms for near-term quantum devices, owing to its hybrid quantum-classical nature and its ability to solve optimization problems using parameterized quantum circuit (PQC) \cite{Peruzzo2014, Cerezo2021, Nguyen_2022}.

The performance of variational quantum algorithms (VQAs) depends critically on the choice of ansatz. Existing approaches, including hardware-efficient ansatz (HEA) \cite{Kandala2017}, the quantum approximate optimization algorithm (QAOA), and Hamiltonian variational ansatz \cite{PhysRevA.92.042303, PRXQuantum.1.020319}, have demonstrated encouraging results for quantum chemistry and combinatorial optimization. However, designing an ansatz that simultaneously provides sufficient expressibility to contain the target state and favorable trainability remains a challenge. Highly expressive circuits are capable of representing a large class of quantum states, but they often suffer from barren plateaus, where gradients vanish exponentially with the system size and optimization becomes prohibitively difficult \cite{Wang2021}.

Among the various circuit families proposed for near-term quantum computing, Instantaneous Quantum Polynomial-time (IQP) circuits constitute a particularly interesting class. Introduced by Shepherd and Bremner \cite{10.1098/rspa.2008.0443}, IQP circuits consist entirely of commuting gates and have attracted considerable attention due to the conjectured classical hardness of their sampling problem \cite{Bremner2016, rajakumar2025polynomial, Bremner2017}. Unlike generic HEA, IQP circuits possess a diagonal structure that naturally generates Pauli-$Z$ correlations and can be implemented with shallow circuit depth and a high degree of parallelism \cite{Bremner2017, bremner_classical2010, Paletta2024robustsparseiqp}. These features make IQP circuits attractive candidates for near-term quantum processors, where coherence times and gate fidelities remain limited \cite{placidi2026impactqubitconnectivityquantum}. Furthermore, recent developments have extended IQP circuits beyond sampling applications to quantum generative models and variational learning tasks \cite{Benedetti_2019, Ball_Gimbernat_2026}.

It is important to distinguish IQP circuits from the broader VQE framework. VQE specifies a hybrid procedure in which a parameterized quantum state is prepared, its energy is measured, and the parameters are updated by a classical optimizer; it does not prescribe a unique circuit architecture \cite{Peruzzo2014}. HEA, Hamiltonian variational ansätze, QAOA, and the IQP circuits considered here are therefore alternative ansatz choices within this general variational paradigm. IQP ansätze are distinguished by a parameterized unitary generated by mutually commuting Pauli operators, which permits the phase-separation operations to be scheduled in parallel when the hardware connectivity allows and provides direct control over the correlation structure through the interaction graph \cite{10.1098/rspa.2008.0443, Bremner2017, bremner_classical2010}. For diagonal Ising- and QUBO-type Hamiltonians, their $Z$-diagonal generators also align naturally with the one- and two-body terms of the objective Hamiltonian. Compared with generic HEA circuits, this structure is more constrained but also more interpretable: connectivity determines which correlations can be represented, while avoiding unnecessary connectivity may reduce circuit depth, routing overhead, and gradient concentration. Compared with QAOA, whose cost unitary is fixed by the problem Hamiltonian and alternated with a noncommuting mixer, a variational IQP ansatz allows its interaction graph and individual commuting phases to be selected independently of the target Hamiltonian \cite{ZHANG2026129002}. These properties do not imply that IQP circuits are universally superior to VQE or QAOA. Rather, they make IQP a structured ansatz family in which implementation cost, global expressibility, and trainability can be related directly to circuit connectivity. Establishing when that structure is beneficial is precisely the purpose of the present study.

IQP circuits have recently been employed as variational ansatze for combinatorial optimization, either as an extension of 1-layer QAOA with classically computed optimal parameters \cite{PhysRevResearch.6.013071} or within the filtering VQE \cite{Marin-Sanchez_2025}. These studies, however, each fix a single circuit architecture within a specific protocol, and the role of connectivity in the trainability of IQP ansatze has not been examined. Existing studies have primarily focused on the computational complexity of IQP sampling \cite{Bremner2016,Bremner2017,bremner_classical2010} or on their application as quantum circuit Born machines. Comparatively little is known about their trainability, expressivity, and optimization capabilities when employed as variational ansatz. In particular, the relationship between circuit connectivity, entanglement generation, barren plateaus, and optimization performance has not yet been systematically investigated. Understanding this relationship is crucial for determining whether IQP circuits can provide practical advantages beyond sampling tasks and for identifying suitable architectures for large-scale Hamiltonian optimization.

In this work, we present a comprehensive study of IQP circuits for Hamiltonian optimization. We consider three representative circuit architectures with increasing connectivity, namely single-$Z$, circular, and fully connected IQP circuits, and evaluate their performance on three benchmark Hamiltonians corresponding to the classical Ising model, MaxCut, and Number Partition problems. To place the IQP results in context, we additionally report the behaviour of two standard variational ansatz, the HEA and the QAOA; these belong to different circuit families and serve as qualitative reference points rather than benchmarking targets. To characterize the optimization behavior of these architectures, we analyze their relative approximation ratios and gradient variances under different parameter initialization strategies. In addition, we investigate the expressibility and bipartite entanglement entropy of each circuit structure in order to understand the connection between representational power and trainability. The main contributions of this work are summarized as follows:

\begin{enumerate}
\item We provide a systematic evaluation of IQP circuits as variational ansatz for Hamiltonian optimization across the classical Ising, MaxCut, and Number Partition problems.
\item We analyze the expressibility and entanglement properties of different IQP architectures and investigate how these properties correlate with optimization performance.
\item We characterize the scaling behavior of gradient variance and demonstrate how increasing connectivity leads to stronger barren plateau effects.
\item We classify five parameter initialization strategies into full-range and small-variance classes and show how the connectivity-trainability trade-off differs across the two classes.
\end{enumerate}

\section{Background and theoretical framework}
\label{sec:background}
\subsection{Instantaneous Quantum Polynomial-time circuit}
\label{subsec:iqp}
The Instantaneous Quantum Polynomial-time (IQP) circuit on an $n$-qubit system consists of three stages \cite{ placidi2026impactqubitconnectivityquantum, armengol2026iqpoptfastoptimizationinstantaneous, lerch2026iqpbornmachinesdatadependent}. Starting from the initial state  $|\bm 0\rangle \equiv |0 \rangle^{\otimes n}$, a layer of Hadamard gates is first applied to prepare a uniform superposition. This is followed by a diagonal unitary $D(\bm\theta)$. A second layer of Hadamard gates is then applied before measurement in the computational basis. The overall unitary transformation implemented by the IQP circuit is given by

\begin{align}
U(\bm\theta)|\bm 0\rangle = H^{\otimes n}D(\bm\theta)|\bm+\rangle.
\end{align}

\begin{figure}[t]
    \centering
    \includegraphics[width=\textwidth]{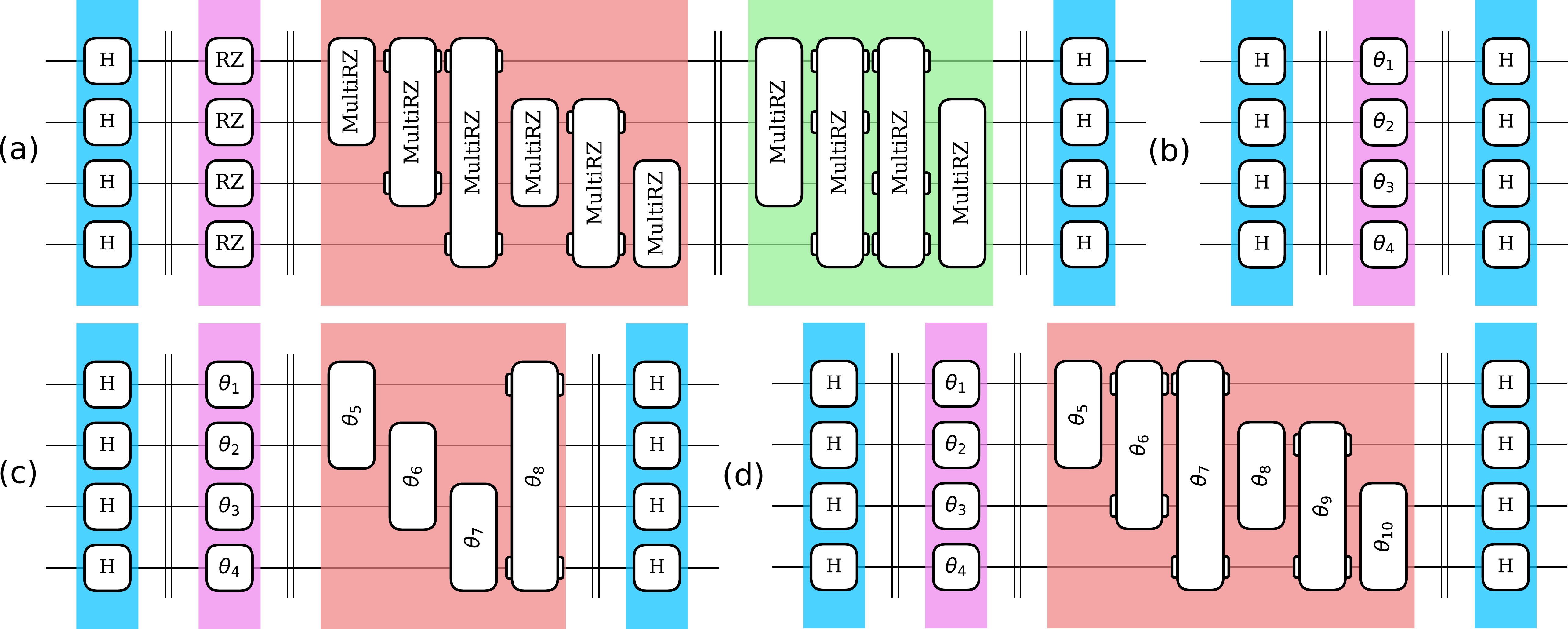}
    \caption{Structure of an IQP circuit where $Z_{j}$ denotes a tensor product of Pauli-$Z$. (a) General IQP (single-Z + ZZ + ZZZ), (b) Single-Z consists solely of single-qubit Pauli-$Z$ rotations $D_S(\bm{\theta})=\exp\left(i\sum_j \theta_j Z_{j}\right)$ (c) Circular connectivity $D_{CC}(\bm{\theta}) = \exp\left( i\sum_j \theta_j Z_j + i\sum_{k} \theta_{k, k+1} Z_k Z_{k+1} \right)$ where two-qubit interactions are arranged in a ring topology, this architecture captures local pairwise correlations while maintaining a sparse interaction pattern. (d) Full connectivity $D_{FC}(\bm{\theta}) = \exp\left( i\sum_j \theta_j Z_j + i\sum_{i<j}\theta_{ij}Z_iZ_j \right)$, where every pair of qubits is coupled through a two-body interaction.}
    \label{fig:iqp_circuit}
\end{figure}

In this work, we consider three representative IQP circuit architectures (single-$Z$, circular connectivity, full connectivity) with different levels of connectivity in order to investigate the interplay between circuit expressivity and optimization performance, illustrated in Fig.~\ref{fig:iqp_circuit}. Compared with single-$Z$ and circular connectivity, the fully connected structure possesses the largest number of variational parameters and the highest global expressibility.

\begin{figure}[t]
    \centering
    \includegraphics[width=\textwidth]{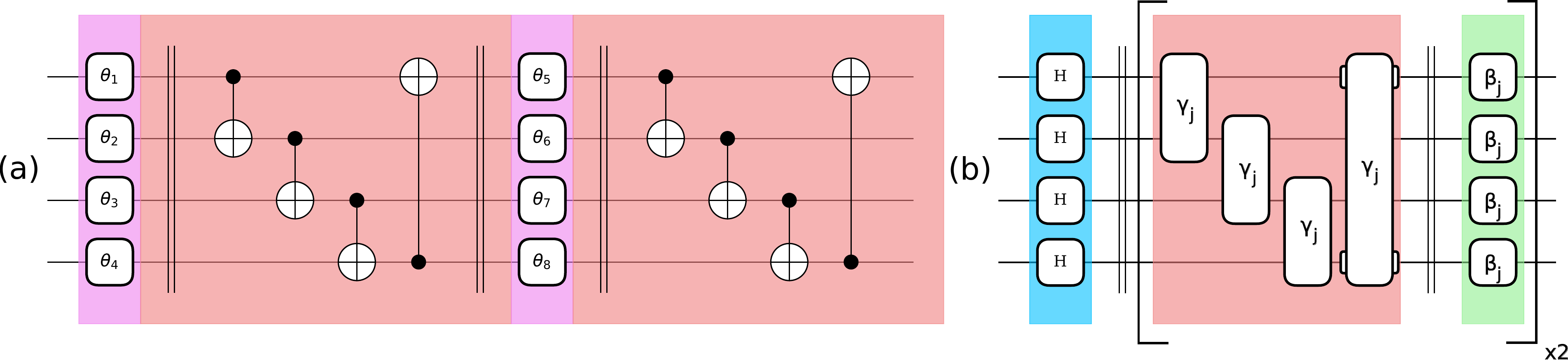}
    \caption{Reference ansatz. (a) HEA $U_{\mathrm{HEA}}(\bm\theta)=\prod_{l=1}^{L}\Bigl[W\prod_{j=1}^{n}R_Y^{(j)}(\theta_{l,j})\Bigr]$ with $L\times n$ parameters and $L=2$ layers: each composed of a single-qubit rotation on every qubit followed by an entangling layer of CX gates arranged in a ring topology. (b) QAOA for a
    representative MaxCut instance. The QAOA ansatz prepares the state $|\psi_p(\bm\gamma, \bm\beta)\rangle=\prod_{l=1}^{p}e^{-i\beta_l H_M}\,e^{-i\gamma_l H_C}\,|{+}\rangle^{\otimes n}$, where $H_C$ is the target Hamiltonian, $H_M = \sum_{j=1}^{n} X_j$ is the transverse-field mixer, and $(\bm\gamma, \bm\beta)$ are $2p$ parameters ($p-2$).}
    \label{fig:ref_ansatz}
\end{figure}

As presented in Fig.~\ref{fig:ref_ansatz} (a), each HEA layer consists of a rotation $R_Y(\theta)=e^{-i\theta Y/2}$ on every qubit followed by the entangling layer $W=\mathrm{CX}_{n,1}\,\mathrm{CX}_{n-1,n}\cdots\mathrm{CX}_{2,3}\,\mathrm{CX}_{1,2}$. The CX gates act sequentially, starting from $\mathrm{CX}_{1,2}$, and $\mathrm{CX}_{n,1}$ closes the ring; $W$ is therefore a cascade of CX gates around a ring rather than a brick-wall layer. In a general HEA, the entangling layer follows the full coupling graph of the device~\cite{Kandala2017}; our choice of a ring corresponds to a device whose coupling graph contains an $n$-qubit cycle.
\subsection{Metrics}

Following Sim et al. \cite{exp}, the expressibility is quantified by comparing the distribution of fidelities $x \equiv |\langle \phi | \psi \rangle|^{2}$ between pairs of independently sampled circuit states with that of Haar-random states $f(x) = (d -1)(1-x)^{d-2}$ with $d=2^n$, which is shown in Theorem 1 and derived in Appendix~\ref{appendix:theorem1}.

Then, the expressibility of the IQP circuit is the Jensen-Shannon distance \cite{MENENDEZ1997307, 61115} between the Haar-random distribution and the output distribution of the IQP circuit as:
\begin{align}
    d_{JS} = \sqrt{\frac{D_{KL}(p || m) + D_{KL}(q || m)}{2}},
\end{align}
where $m$ is the pointwise mean of $p$ (Haar-random distribution) and $q$ (output distribution of the IQP circuit), and $D$ is the Kullback-Leibler divergence. We note that $d_{JS}$ quantifies global expressibility. This differs from the ability of the ansatz to represent a particular target state. Throughout this work, ``expressibility'' refers to the global notion.


The entanglement entropy links global expressibility and trainability. It is a prerequisite for global expressibility: Haar-typical states are near-maximally entangled \cite{PhysRevLett.71.1291}, so reproducing the Haar ensemble requires generating entanglement. Excess entanglement, however, suppresses trainability, since volume-law entanglement induces barren plateaus \cite{PRXQuantum.2.040316}. It is therefore the common physical origin of the connectivity-trainability trade-off.

For a pure state $ \left| \psi \right> $ \cite{Janzing2009}, divide the qubits into two subsystems $ A \cup B = \{ 1,\ldots, n \} $. For this study, we choose a half-half bipartition where $ \left| A \right| = \lfloor n/2 \rfloor $. The reduced density matrix is $\rho_{A} = \operatorname{Tr_{B}}(\left| \psi \right> \left< \psi \right|)$. The von Neumann entanglement entropy is then defined as

\begin{align}
    S(\rho_{A}) = - \operatorname{Tr}(\rho_{A} \log_{2} \rho_{A}).
\end{align}

\subsection{Barren Plateaus}
\label{subsec:bp}

The VQAs optimize a PQC by minimizing a cost function through iterative updates of the circuit parameters \cite{Cerezo2021VQA}. The effectiveness of this procedure relies on the availability of sufficiently large gradients that provide useful information for optimization. However, it has been shown that many PQCs suffer from the phenomenon known as \emph{barren plateaus}, where the gradients of the cost function vanish with the number of qubits \cite{McClean2018}.

Consider a parameterized quantum state $|\psi(\bm{\theta})\rangle
=
U(\bm{\theta})
|\bm 0\rangle$, where $U(\bm{\theta}) $ is a PQC with
$\bm{\theta}=(\theta_1,\theta_2,\ldots,\theta_m)$. Given a target Hamiltonian $\hat{H}$, the objective is to minimize the expectation value $\mathcal{C}(\bm{\theta})=\langle \psi(\bm{\theta}) |H|\psi(\bm{\theta}) \rangle$, which serves as the cost function. Training is typically performed using gradient-based optimization methods. The gradient with respect to parameter $\theta_k$ is given by $\partial \mathcal{C}(\bm{\theta})/\partial \theta_k$ and the trainability of the circuit is characterized by the variance of the gradient $\mathrm{Var}
\left[
\partial \mathcal{C}/\partial \theta_k
\right]$. A barren plateau occurs when the gradient variance decreases,

\begin{align}
\mathrm{Var}
\left[
\partial \mathcal{C}/\partial \theta_k
\right]
=
\mathcal{O}(b^{-n}),\;(b>1).
\end{align}

From a practical perspective, barren plateaus imply that an exponentially large number of measurements is required to estimate the gradients with a fixed accuracy. As a result, parameter updates become inefficient, and the circuit may fail to converge to low-energy states. The gradient variance, therefore, serves as an important indicator of trainability. Larger gradient variances correspond to steeper optimization landscapes and more effective parameter updates, whereas rapidly decaying variances signal the onset of barren plateaus.

\section{Experimental setup}
\label{sec:results}

The experiments are divided into two parts. The first investigates optimization performance and trainability through the relative approximation ratio and the gradient variance; the second analyzes the expressibility and entanglement properties of all circuits.

\subsection{Benchmarking problems}
\label{subsec:hamiltonian}

To provide a comprehensive evaluation, we consider three Hamiltonians with distinct structures and interaction patterns: the Ising Hamiltonian, the MaxCut Hamiltonian, and the Number Partition Hamiltonian. These benchmark problems encompass local fields, sparse graph interactions, and dense all-to-all couplings, thereby allowing us to study how the connectivity of the target Hamiltonian influences the optimization capability of IQP circuits. For each problem and each system size $n \in \{3,6,9,12,15\}$ we use $50$ instances; every ansatz and initialization strategy is evaluated on the same instances.

The first benchmark is the classical Ising Hamiltonian \cite{gallavotti1999statistical},

\begin{align}
H_{\mathrm{Ising}} = \sum_i h_i Z_i + \sum_{i<j} J_{ij} Z_iZ_j,
\end{align}
where $h_i$ denotes the local field acting on qubit $i$, and $J_{ij} $ represents the coupling strength between qubits $i$ and $j$. $\{h_i,J_{ij}\}$ are generated following uniform distribution $\mathcal{U}(0,1)$. The Ising model plays a central role in statistical mechanics and serves as a generic framework for expressing a wide range of quadratic unconstrained binary optimization (QUBO) problems. Its ground state corresponds to the spin configuration that minimizes the total energy determined by both local fields and pairwise interactions. The exact ground-state energy is obtained by the brute-force method.

The second benchmark is the MaxCut Hamiltonian \cite{PhysRevA.97.022304},
\begin{align}
H_{\mathrm{MaxCut}} = - \sum_{(i,j)\in E(G)} w_{ij} \frac{I-Z_iZ_j}{2},
\end{align}
where $G=(V,E)$ is a random weighted graph on $n$ vertices (one per qubit) in the Erd\H{o}s--R\'enyi family \cite{ErdosRenyi1959} and $w_{ij}$ denotes the weight associated with edge $(i,j)$. Each of the $n(n-1)/2$ possible edges is included independently with probability $p=3/n$ \cite{Gilbert1959}. Each edge $(i,j)$ carries an integer weight $w_{ij}$ drawn uniformly from $\{1,\dots,10\}$. The MaxCut problem seeks a partition of the vertices into two disjoint subsets such that the total weight of the edges crossing the partition is maximized. By encoding the graph into a Hamiltonian, each edge contributes energy when the corresponding spins are assigned opposite signs. The exact solution is also found by the brute-force method. MaxCut is one of the most widely studied NP-hard problems and has become a standard benchmark for quantum optimization algorithms.

The third benchmark is the Number Partition Hamiltonian \cite{PhysRevLett.81.4281},
\begin{align}
H_{\mathrm{Partition}} = \left( \sum_{i} a_i Z_i \right)^{2},
\end{align}
where $a_i$ are drawn uniformly at random from $\{1,\dots,99\}$, and the whole vector $(a_1,\dots,a_n)$ is redrawn until the Karmarkar-Karp differencing heuristic \cite{Boettcher2008} finds a perfect partition. The number partitioning problem aims to divide the integers into two groups whose sums are as nearly equal as possible. The corresponding Hamiltonian penalizes imbalances between the two subsets, and its ground state represents the partition with the minimum difference in total weight. Owing to its dense interaction structure, this problem provides a challenging benchmark for assessing the expressivity of variational quantum circuits.

\subsection{Optimization experiments}

Simulations are performed with exact state-vector evaluation (PennyLane \texttt{lightning.qubit}, \texttt{shots=None}), following the IQP formulation of the \texttt{IQPopt} package \cite{armengol2026iqpoptfastoptimizationinstantaneous}. Each optimization run consists of 2500 iterations of the Adam optimizer with a step size of $10^{-3}$, and every configuration is averaged over 50 above instances.

\begin{align}
r_{RA}=\left|
\frac{f(\bm\theta)-f(\bm\theta^*)}{f(\bm\theta^*)}
\right|,
\end{align}
where $f(\bm\theta)$ denotes the objective value obtained by the IQP circuit and $f(\bm\theta^*)$ is the exact optimal value. Smaller approximation ratios correspond to better optimization performance. We report $r_{RA}$ on the unnormalized Hamiltonians to allow direct comparison with existing studies.

\subsection{Expressibility and entanglement properties}

For each connectivity structure, circuit parameters are sampled independently from a uniform distribution over the interval $(0,\pi)$. The resulting quantum states are extracted from the simulator and used to evaluate the expressibility and entanglement entropy. Expressibility is quantified by comparing the distribution of pairwise state fidelities with the corresponding Haar-random distribution, while the entanglement entropy provides a measure of the correlations generated by the circuit. Together, these metrics offer insight into the global expressibility of the different IQP architectures and their relationship with optimization performance and trainability.

\subsection{Parameter initialization}

To investigate the effect of parameter initialization, five different initialization strategies are employed for each connectivity structure (Gaussian $\mathcal{N}(0,1)$, uniform $\mathcal{U}(-\pi,\pi)$, near-$\pi/4$, Gaussian LeCun, and Gaussian He), as summarized in Table~\ref{tab:init_methods}. These strategies are motivated by previous observations that suitable parameter concentration can improve trainability and mitigate barren plateaus \cite{Grant2019initialization, 3600270.3601622, PhysRevApplied.22.054005}.

The last two strategies are the quantum heuristics obtained by Peng \emph{et al.} \cite{11250281}, by identifying the fan-in of a layer with the total number of variational parameters $n_{\mathrm{params}}$ of the circuit. Both carry a scale factor $\gamma$, which we fix as in \cite{3600270.3601622},
\begin{align}
\label{eq:gamma}
\gamma^{2} = \frac{1}{4S(K+2)},
\end{align}
where $S$ is the largest number of non-identity Pauli operators carried by a single term of the target Hamiltonian, and $K$ is the depth of the ansatz. All three benchmark Hamiltonians of Sec.~\ref{subsec:hamiltonian} are at most two-local, so $S=2$ in this work and $\gamma^{2}$ varies only through $K$. We take $K$ to be  the depth of the undecomposed ansatz, the longest path through its gate graph; it is fixed by the architecture and $n$, except for the QAOA, whose cost layer follows the instance Hamiltonian.

\begin{table}[t]
\centering
\caption{Parameter initialization strategies used in this work.\label{tab:init_methods}}
\begin{tabular*}{\textwidth}{@{\extracolsep\fill}p{0.25\textwidth}p{0.25\textwidth}p{0.40\textwidth}@{\extracolsep\fill}}
\toprule
\textbf{Strategy} & \textbf{Formula} ($\theta_i \sim$) & \textbf{Description} \\
\midrule
Gaussian Initialization \cite{3600270.3601622} &
$\mathcal{N}(0,1)$ &
Parameters are sampled independently from a standard Gaussian distribution. \\
Uniform Initialization \cite{McClean2018} &
$\mathcal{U}(-\pi,\pi)$ &
Parameters are sampled uniformly. \\
Near-$\pi/4$ Initialization \cite{PhysRevApplied.22.054005} &
$\mathcal{U}\!\left(\frac{\pi}{4}-\epsilon,\frac{\pi}{4}+\epsilon\right)$ &
Parameters are initialized $\approx\pi/4$ with small random perturbations controlled by $\epsilon$. \\
Gaussian LeCun Initialization \cite{11250281} &
$\mathcal{N}\!\left(0,\, \gamma^2\cdot\frac{1}{n_{\mathrm{params}}}\right)$ &
Parameters are sampled from a Gaussian whose variance decreases with the number of parameters, scaled by $\gamma^2$. \\
Gaussian He Initialization \cite{11250281} &
$\mathcal{N}\!\left(0,\, \gamma^2\cdot\frac{2}{n_{\mathrm{params}}}\right)$ &
Parameters are sampled as in LeCun, with twice the variance.\\
\bottomrule
\end{tabular*}
\end{table}

We group five strategies into two classes according to the variance of the initial parameter distribution. \emph{Full-range initializations} --- the Gaussian $\mathcal{N}(0,1)$ and the uniform $\mathcal{U}(-\pi,\pi)$ distribution, have variance of order unity, so that the initial parameters spread over the full period of the rotation angles; this is the regime in which barren-plateau results for random initialization apply \emph{Small-variance initializations} --- near-$\pi/4$, He, and LeCun, concentrate the parameters around a fixed point with a variance that is either small ($\epsilon = 0.05$ for near-$\pi/4$) or decreases with the system size ($\gamma^2\cdot \frac{2}{n_{\mathrm{params}}}$ and $\gamma^2\cdot \frac{1}{n_{\mathrm{params}}}$, respectively), following the observation that reduced-variance initialization can mitigate barren plateaus.

\section{Results}
\label{subsec:results}

\subsection{Optimization performance}
\label{subsec:ratio}

\begin{figure}[t]
    \centering
    \includegraphics[width=\textwidth]{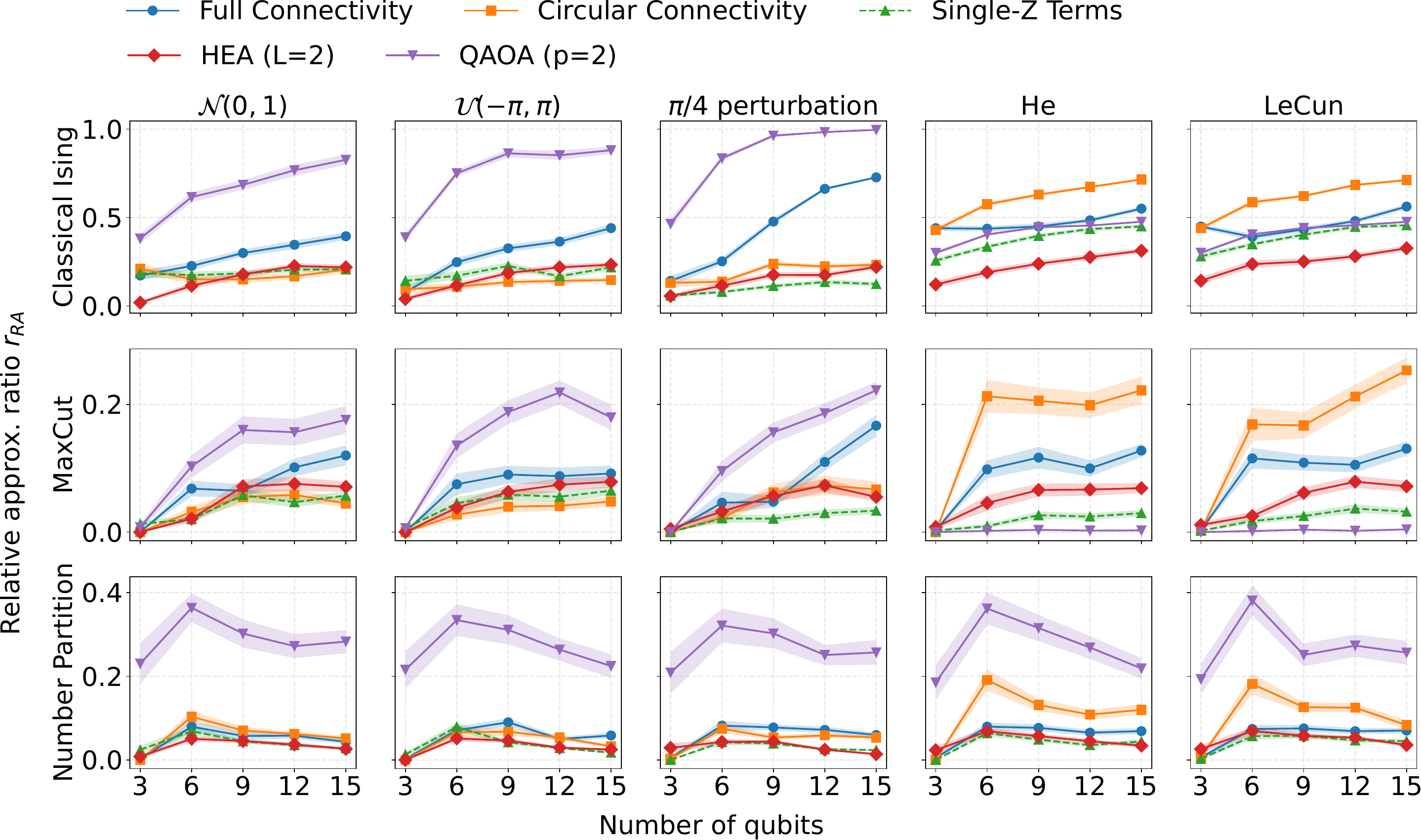}
    \caption{Comparison of the optimization error $r_{RA}$ achieved for three Hamiltonians (rows) and five parameter initialization strategies (columns). Lower is better.}
    \label{fig:ratio}
\end{figure}

Fig.~\ref{fig:ratio} shows $r_{RA}$ for the three Hamiltonians and the five initialization strategies. Within the IQP family, the fully connected circuit is never the best IQP architecture for $n \geq 6$ in any combination of problem and initialization. Under the two full-range initializations, the sparse circuits keep an almost size-independent ratio on all three problems, while the ratio of the fully connected circuit grows steadily with $n$, more than doubling on Ising between $n=3$ and $n=15$. The three small-variance initializations do not act uniformly on the IQP family. The near-$\pi/4$ initialization favours the sparse circuits, giving the single-$Z$ circuit its lowest ratios on Ising and MaxCut, but pushes the fully connected circuit in the opposite direction, up to 0.73 on Ising. The He and LeCun initializations, whose variance decreases with the system size, instead penalize the circular circuit: on all three problems it ends with the largest $r_{RA}$ of the IQP architectures, while the single-$Z$ circuit keeps the smallest.
The two reference ansätze frame these results. The HEA is the least initialization-sensitive ansatz in the comparison, and under the full-range initializations its ratios are comparable to the sparse IQP circuits. The QAOA at $p=2$ ends well above every other ansatz on Number Partition, and on Ising except under He and LeCun; with only two parameters per layer, and a cost layer tied to the problem Hamiltonian, it has little capability for optimization at this depth. Its behaviour on MaxCut is the most initialization-dependent of any ansatz we test: it remains the worst under the full-range and near-$\pi/4$ initializations, but the best under He and LeCun.

\subsection{Trainability}
\label{subsec:gv}

\begin{figure}[htbp]
    \centering
    \includegraphics[width=\textwidth]{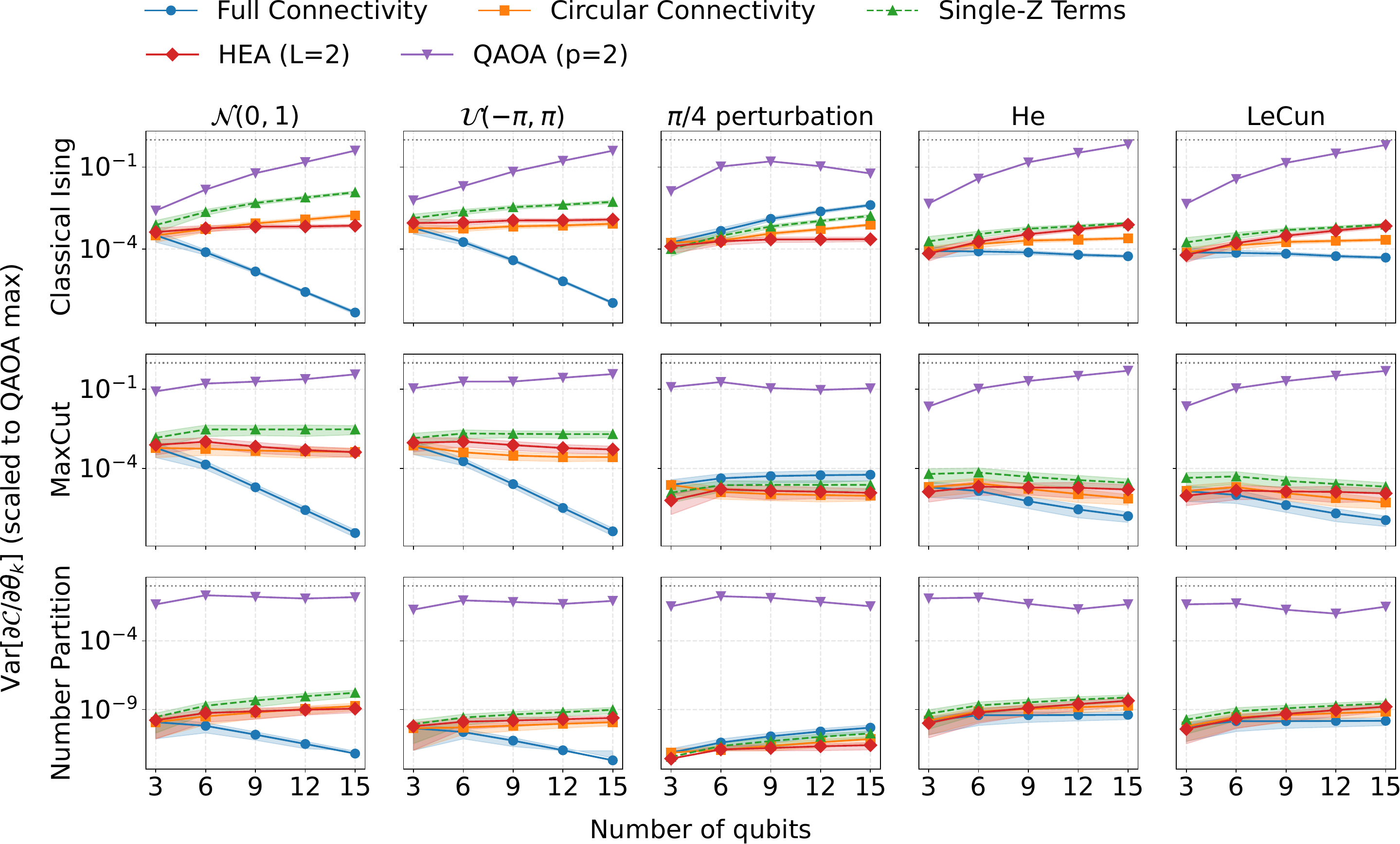}
    \caption{Gradient variance $\mathrm{Var}[\partial\mathcal{C}/\partial\theta_k]$ with the same setup as in Fig.~\ref{fig:ratio}. Each curve is normalized to the QAOA maximum within its panel, and shaded bands show one standard deviation over instances (QAOA shown as mean only). }
    \label{fig:gradient}
\end{figure}

The trainability of the different circuits is examined in Fig.~\ref{fig:gradient} through the scaling of the gradient variance with the number of qubits. Under the two full-range initializations, the fully connected circuit is the only IQP architecture whose variance decays exponentially: on the Ising problem it falls by nearly three orders of magnitude between $n=3$ and $n=15$, the signature of a barren plateau. The circular, single-$Z$, and HEA circuits show no exponential decay: their variance grows with $n$ on Ising and Number Partition and changes only mildly on MaxCut. Consistent with Fig.~\ref{fig:ratio}, this accounts for the poor full-range performance of the fully connected circuit: a larger parameter space does not help when the gradient signal needed to explore it vanishes. The QAOA curves lie above the rest because a single $\gamma_l$ multiplies the entire cost Hamiltonian, whereas each IQP and HEA parameter generates a single Pauli term; the scale of $\partial\mathcal{C}/\partial\gamma_l$ is therefore set by $H_C$ as a whole rather than by one of its terms. The gap is largest for Number Partition, whose Hamiltonian has the largest norm of the three. It reflects the scale of the QAOA parameters rather than their trainability.

The small-variance initializations remove the exponential decay, but they do not act alike. Under near-$\pi/4$ initialization the ordering by connectivity inverts: the variance of the fully connected circuit grows with $n$ and ends above the sparse circuits. Under He and LeCun the ordering is preserved, the fully connected circuit remaining the smallest at every size, but without the collapse. The initialization class therefore decides whether the fully connected circuit starts on a barren plateau.

\subsection{Expressibility and entanglement}

\label{subsec:expr}

\begin{figure}[htbp]
    \centering
    \includegraphics[width=0.99\textwidth]{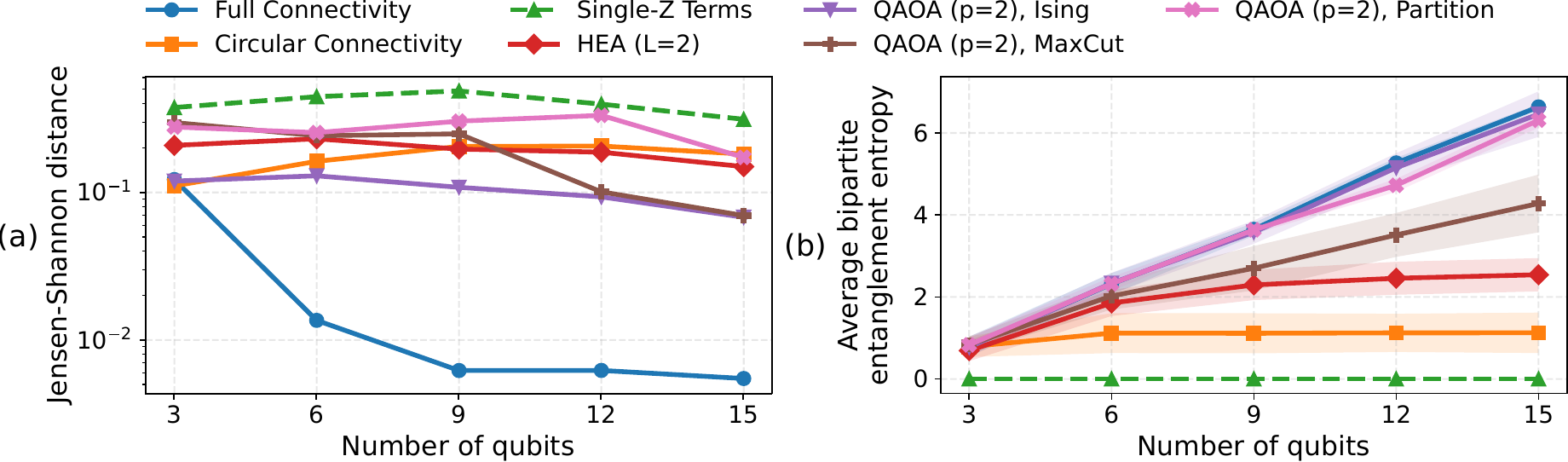}
    \caption{Scaling behavior of the five ansatze with respect to the number of qubits. }
    \label{fig:expr}
\end{figure}

To trace the trade-off back to the circuit properties, Fig.~\ref{fig:expr} shows the expressibility and the bipartite entanglement entropy of the five ansatz, with the problem-dependent QAOA evaluated separately for each Hamiltonian. In Fig.~\ref{fig:expr}(a), the Jensen-Shannon distance to the Haar distribution decreases with connectivity at almost every system size: at $n=15$ the fully connected circuit reaches $d_{JS}=5.5\times10^{-3}$, an almost Haar-random output ensemble, while the circular and single-$Z$ circuits remain at 0.18 and 0.31, with the HEA below both at 0.15. The entanglement entropy in Fig.~\ref{fig:expr}(b) separates the architectures more sharply. The entropy of the fully connected circuit grows almost linearly with $n$ and reaches $S=6.64$ at $n=15$, close to the maximal value $\lfloor n/2 \rfloor = 7$ and consistent with volume-law scaling. The circular circuit saturates near $S \approx 1.1$ independent of the system size, and the single-$Z$ circuit generates no entanglement at all, since its diagonal layer contains only single-qubit terms and the output remains a product state.

Taken together, these results indicate that neither maximum global expressibility nor maximum trainability alone governs optimization performance. Fully connected circuits generate high global expressibility and strongly entangled states, but their gradient variance decays with system size, hindering optimization and yielding the worst approximation ratios among the IQP architectures. The circular and single-$Z$ circuits, despite differing sharply in entanglement, both maintain stable gradients and reach comparable, substantially lower errors, indicating that entanglement is not required to represent the ground states of these diagonal Hamiltonians. Circular connectivity is the minimal architecture that retains entanglement capacity while preserving trainability; we therefore adopt it as the balanced default among the IQP ansatz, with single-$Z$ equally effective on the present benchmarks but generating no entanglement. Table~\ref{tab:summary} summarizes these qualitative trends.

\begin{table}[t]
\centering
\caption{Qualitative comparison of the three IQP architectures, with the HEA and QAOA included for context.\label{tab:summary}}
\begin{threeparttable}
\begin{tabular*}{\columnwidth}{@{\extracolsep\fill}lllll@{\extracolsep\fill}}
\toprule
\textbf{Ansatz} & \textbf{Global Expr.} & \textbf{Entangl.} & \textbf{GV} & \textbf{Perf.} \\
\midrule
Full        & High        & Volume-law & Decays rapidly  & Degrades \\
Circular    & Moderate    & Saturating & Stable          & Strong   \\
Single-$Z$  & Low         & None       & Stable          & Strong   \\
\midrule
HEA$^\ddagger$  & Mod.-High & Saturating & Stable            & Strong \\
QAOA$^\ddagger$ & High       & High       & N/A$^\dagger$     & Poor \\
\bottomrule
\end{tabular*}
\begin{tablenotes}
\item[$^\dagger$] is not comparable across families: the QAOA generator
is the problem Hamiltonian, so its variance scales with $\lVert H\rVert^2$.
\item[$^\ddagger$] Reference ansatz, shown for context rather than benchmarking.
\end{tablenotes}
\end{threeparttable}
\end{table}

\section{Discussion}
\label{sec:discuss}

Increasing circuit connectivity substantially enhances global expressivity and entanglement generation, with fully connected IQP circuits approaching Haar-like behavior and nearly volume-law entanglement scaling. In our experiments, this is consistently associated with rapidly decaying gradient variances: the fully connected circuit is the only IQP architecture whose gradient variance decays exponentially, and also the one with the highest entanglement entropy, across all three Hamiltonians and both full-range initializations. The pattern is nonetheless consistent with the mechanism proposed in prior work, whereby high, volume-law entanglement concentrates local expectation values around their global average and suppresses gradient magnitude \cite{PRXQuantum.2.040316}. Under this reading, the global expressibility of highly connected circuits does not translate into better optimization here: the sparse architectures keep stable gradients and generally reach lower errors, the main exception being He and LeCun initializations, under which the circular circuit degrades. These findings suggest IQP circuit effectiveness is governed by a balance between global expressivity and trainability rather than maximum expressivity alone.

The single-$Z$ circuit is particularly informative: although it generates no entanglement, it matches the more connected circuits on every benchmark. This is expected once we note that the ground state of a diagonal Hamiltonian is a computational basis state, which a product ansatz already represents exactly, so that expressivity beyond single-qubit rotations brings no benefit while its cost in trainability is real. The limitation of single-$Z$ is therefore one of scope rather than performance: lacking entanglement, it cannot extend to target Hamiltonians whose ground states are themselves entangled. Circular connectivity occupies the intermediate regime that we recommend as a default, combining stable trainability with the entanglement capacity needed for such more general targets.

\begin{figure}[t]
    \centering
    \includegraphics[width=0.99\textwidth]{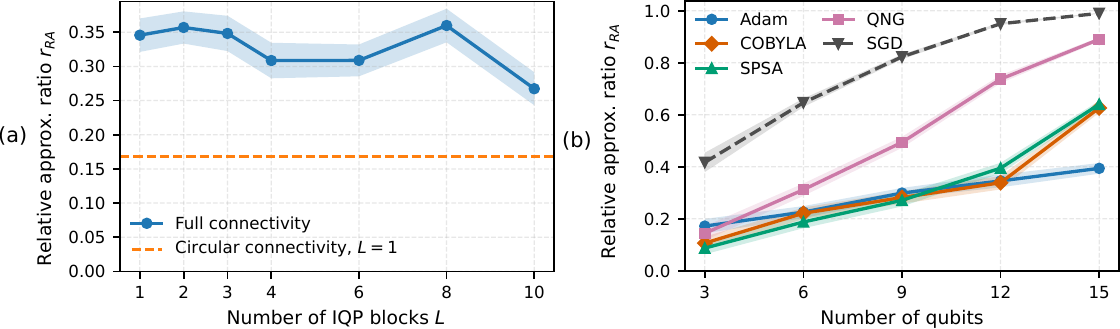}
    \caption{Robustness of the connectivity-trainability trade-off, on the fully
connected IQP circuit. (a) $r_{RA}$ against the number of stacked IQP blocks
$L$ at $n=12$. (b) $r_{RA}$ against system size for five optimizers at $L=1$.
Shaded bands show one standard error of the mean; the two panels do not share
the vertical scale.}
    \label{fig:robust}
\end{figure}

The trade-off established above was measured at a single circuit depth and with a single optimizer, which leaves open whether it is a property of the connectivity or a consequence of that training configuration. We therefore repeat the measurement along these two axes. Both checks use the same configuration: the fully connected circuit on the Ising Hamiltonian with $\mathcal{N}(0,1)$ initialization, averaged over 50 instances per point. Every run is given the same budget of 2500 iterations, and the gradient-based optimizers share the same step size of $10^{-3}$; COBYLA and SPSA have no comparable step-size parameter and are run at their library defaults. We stack $L$ IQP blocks, each carrying its own parameters, and vary layers $L = [1, 2, 3, 4, 6, 8, 10]$ at $n=12$ qubits. Each block repeats the structure of Sec.~\ref{subsec:iqp}: a Hadamard layer, the diagonal unitary $D(\bm\theta_l)$ of the fully connected architecture, and a second Hadamard layer. Fig.~\ref{fig:robust} (a) shows that $r_{RA}$ is insensitive to $L$. It stays between $0.27$ and $0.36$ over the whole range, moving up and down rather than in one direction, with no monotone trend; the spread across $L$ is comparable to the instance-to-instance error. The curve also never approaches the dashed reference line: ten stacked blocks of full connectivity still end above a single block of circular connectivity at the same system size, problem and initialization. Adding depth does not recover the performance lost to connectivity. We repeat the sweep over $n$ with five optimizers spanning the standard families: gradient descent, Adam, the quantum natural gradient (QNG), the gradient-free Constrained Optimization BY Linear Approximation (COBYLA), and Simultaneous Perturbation Stochastic Approximation (SPSA). In Fig.~\ref{fig:robust} (b), the ranking is size-dependent: at $n=3$ the gradient-free methods reach lower $r_{RA}$ than Adam, by $n=12$ the two are indistinguishable, and at $n=15$ Adam is ahead of every alternative. What does not change with the optimizer is the trend. Every method degrades as $n$ grows, and Adam degrades the slowest, so the optimizer adopted for the main experiments is the one most favorable to the fully connected circuit; any other choice would have strengthened, not weakened, the degradation we report.

\section{Conclusion and future works}

Compared with conventional variational ansatz, IQP circuits possess several attractive properties for Hamiltonian optimization. Their diagonal structure naturally generates Pauli-Z correlations, making them well suited for Ising-type Hamiltonians and QUBO formulations. Furthermore, IQP circuits are shallow, highly parallelizable, and composed entirely of commuting operations, which simplifies both implementation and gradient evaluation on near-term quantum hardware. The present results indicate that these advantages can be effectively exploited when the circuit connectivity is carefully chosen. In particular, moderate connectivity retains entanglement capacity for more general targets while avoiding the severe trainability degradation associated with highly connected architectures. On hardware with restricted connectivity, the circular architecture can be implemented without routing whenever the device coupling graph contains an $n$-qubit cycle. The fully connected architecture requires either all-to-all connectivity or SWAP networks that increase depth and exposure to noise. Hardware constraints would therefore further disadvantage the fully connected circuit.

Three directions for future research follow from these findings. One natural extension is the development of problem-informed IQP architectures whose connectivity reflects the interaction graph of the target Hamiltonian. Such architectures may retain sufficient problem-relevant expressivity while avoiding the severe gradient suppression observed in fully connected circuits. Another promising direction is the incorporation of higher-order Pauli interactions and multi-layer IQP constructions, which may increase global expressibility without requiring all-to-all connectivity. Future work should also investigate techniques for mitigating barren plateaus, including layerwise training, adaptive parameter initialization, local cost functions, and problem-specific optimization strategies.

Beyond Hamiltonian optimization, the connectivity-trainability trade-off identified in this work may have broader implications for VQAs. Similar considerations are likely to arise in quantum machine learning, quantum generative modeling, quantum state preparation, and quantum optimal control, where high global expressibility often faces optimization difficulties. Understanding how circuit structure influences both global expressivity and trainability may therefore provide useful design principles for scalable quantum algorithms more generally.


\roles{Q.C.N. conceived the study, developed the theoretical framework and simulation code, performed the experiments and data analysis, and drafted the manuscript. D.D.T. contributed to the methodology, assisted with validation of the numerical results, and revised the manuscript. V.T.H. supervised the project, secured funding, and revised the manuscript. All authors discussed the results and approved the final version of the manuscript.}

\ack{The authors would like to thank Supanut Thanasilp, Roland Farrell, Marlene Funck, Ricard Puig, and Carla Rieger for helpful discussions. V.T.H. is supported by the VNUHCM - University of Information Technology's Scientific Research Support Fund.}

\section*{Conflicts of Interest}

The authors declare no conflicts of interest.

\data{The code and data supporting the findings of this study are openly available
under the Apache License 2.0~\cite{IQPOptCode}.}


\begin{appendix}
\renewcommand{\theequation}{A\arabic{equation}}

\section{Derivation of the Haar-random fidelity distribution}\label{appendix:theorem1}

\begin{theorem}\label{theorem:1}
    For an $n$-qubit system, the approximate probability density of a single squared amplitude (or fidelity with a fixed pure state) of a Haar-random state is
    \begin{align}
        f(x) = (d -1)(1-x)^{d-2},
    \end{align}
where $d = 2^{n}$, and $x \equiv |\langle \phi | \psi \rangle|^{2} \in [0,1]$ is the fidelity between the Haar-random state $|\psi\rangle$ and a fixed reference state $|\phi\rangle$.
\end{theorem}

To prove this theorem, let $|\psi\rangle = (\psi_{1}, \psi_{2},\ldots, \psi_{d})^{T} $ be a Haar-random pure state in a $d$-dimensional complex Hilbert space, and let $x = |\langle \phi | \psi \rangle|^2$ for some fixed normalized state $|\phi\rangle$. By unitary invariance of the Haar measure, the distribution of $x$ is independent of the choice of $|\phi\rangle$. Therefore, we may choose $|\phi\rangle = \bm e_1 = (1,0,\ldots,0)^T$ so that $x = |\psi_1|^2$. A Haar-random state can be generated as $|\psi\rangle = \frac{g}{\|g\|},$ where $\bm g = (g_1,\ldots,g_d)$ and the $g_i$ are independent standard complex Gaussian random variables ($g_i \sim \mathcal N_{\mathbb C}(0,1)$). We define $Y_i = |g_i|^2$ where $Y_i$ is exponentially distributed with density $f_{Y_i}(y) = e^{-y},\;y \ge 0$. Hence:

\begin{align}
    x = \frac{A}{A+B},\;A = Y_1,\;B = \sum_{i=2}^{d} Y_i
\end{align}

Since $A \sim \Gamma(1,1)$ and $B \sim \Gamma(d-1,1)$. Moreover, $A$ and $B$ are independent, with densities

\begin{align}
    f_A(a) = e^{-a},\;f_B(b) = \frac{b^{d-2} e^{-b}}{\Gamma(d-1)}, \quad b \ge 0.
\end{align}

Introduce the change of variables $x = \frac{A}{A+B}$ and $s = A+B$, so that $A = xs$ and $B = (1-x)s$. The Jacobian is

\begin{align}
\left|
\frac{\partial(A,B)}{\partial(x,s)}
\right|
=
\begin{vmatrix}
s & x \\
-s & 1-x
\end{vmatrix}
=
s
\end{align}

Therefore,
\begin{align} f_{x,s}(x,s) = f_A(xs)\, f_B((1-x)s)\, s \end{align}
Substituting the densities gives
\begin{align}
f_{x,s}(x,s)
=
e^{-xs}
\frac{((1-x)s)^{d-2} e^{-(1-x)s}}
{\Gamma(d-1)}
\, s
\end{align}
Since $e^{-xs}e^{-(1-x)s}=e^{-s}$, we obtain

\begin{align}
    f_{x,s}(x,s)
=
\frac{(1-x)^{d-2}}{\Gamma(d-1)}
s^{d-1}e^{-s}
\end{align}

Integrating out $s$,
\begin{align}
f_x(x)
=
\int_0^\infty f_{x,s}(x,s)\, ds
=
\frac{(1-x)^{d-2}}{\Gamma(d-1)}
\int_0^\infty s^{d-1}e^{-s}\,ds
\end{align}
Using $\int_0^\infty s^{d-1}e^{-s}\,ds = \Gamma(d)$, we find

\begin{align}
    f_x(x) = \frac{\Gamma(d)}{\Gamma(d-1)} (1-x)^{d-2}
\end{align}

Finally, using the identity $\Gamma(d) = (d-1)\Gamma(d-1)$, we conclude that

\begin{align}
    f_x(x) = (d-1)(1-x)^{d-2},\; 0 \le x \le 1
\end{align}

Thus $f(x) = (d-1)(1-x)^{d-2}$ which is the density of the $\mathrm{Beta}(1,d-1)$ distribution.

\end{appendix}

\bibliographystyle{unsrtnat}
\bibliography{refs}

\end{document}